\documentclass[aps,pre,twocolumn,showpacs,superscriptaddress,groupedaddress]{revtex4-2}
\pdfoutput=1
\usepackage{hyperref}
\usepackage{epsfig}
\usepackage{amsmath}
\usepackage{graphicx}
\usepackage{wrapfig}
\usepackage{dcolumn}
\usepackage{bm}
\usepackage{amssymb}
\usepackage{titlesec}
\usepackage{mathtools}
\usepackage{relsize}
\usepackage{cleveref}
\usepackage{calligra}
\usepackage[usenames,dvipsnames]{color}

\begin{document}

\title{Noise-induced peak intensity fluctuations in class B laser systems}
\author{Jason Hindes$^{1}$ and Ira B. Schwartz$^{1}$}
\affiliation{$^{1}$U.S. Naval Research Laboratory, Code 6792,  Washington, DC 20375, USA}

\begin{abstract}
Random perturbations and noise can excite instabilities in population systems that result in large fluctuations. An interesting example involves class B lasers, where the dynamics is determined by the number of carriers and photons in a cavity with noise appearing in the electric-field dynamics. When such lasers are brought above threshold, the field intensity grows away from an unstable equilibrium, exhibiting transient relaxation oscillations with fluctuations due to noise. In this work, we focus on the first peak in the intensity during this transient phase in the presence of noise, and calculate its probability distribution using a Wentzel–Kramers–Brillouin (WKB) approximation. In particular, we show how each value of the first peak is determined by a unique fluctuational-momentum, calculate the peak intensity distribution in the limit where the ratio of photon-to-carrier lifetimes is small, and analyze the behavior of small fluctuations with respect to deterministic theory. Our approach is easily extended to the analysis of transient, noise-induced large fluctuations in general population systems exhibiting relaxation dynamics.     
\end{abstract}
\maketitle

\section{\label{sec:Intro}INTRODUCTION}
Lasers show a rich variety of complex dynamical behaviors caused by nonlinearities, instabilities, and various forms of feedback and coupling architecture including networks of coupled lasers~\cite{LaserDynamics,SemiconductorLasers,RevModPhys.85.421,PrinciplesOfLaserDynamics,shaw2006synchronization}. Important examples of dynamical behavior for laser systems are periodic oscillations, low-frequency fluctuations, spiking, chaos, random number generation, synchronization, nonlinear propagation\cite{schwartz1994subharmonic,shaw2006synchronization,carr2006delayed, billings2004stochastic,Williams2010,PhysRevLett.105.264101,PhysRevLett.118.123901,NonlinearOptics,FundamentalsOfPhotonics}, etc. Because of their fast time scales, precise physical control, and diversity of spatiotemporal behaviors, lasers remain an important platform for studying high dimensional nonlinear systems, including bifurcations and control of turbulence instabilities~\cite{Selmi2016,scholl2001nonlinear,POSTLETHWAITE200765} and novel phase transitions\cite{PhysRevLett.114.063903}. 

%Moreover, they have been instrumental in experimentally and theoretically examining high dimensional bifurcations of turbulent instabilities and their control \cite{Selmi2016,scholl2001nonlinear,POSTLETHWAITE200765}. Because of their fast time scales, precise physical control, and diversity of spatiotemporal behaviors, lasers remain an important platform for studying high dimensional nonlinear systems, such as turbulence in addition to their immense practical use as a source of coherent light\textcolor{red}{[]}.

A widespread category of laser is the class B, in which the dynamics is effectively determined by the number of photons and the number of carriers in a laser cavity\cite{PrinciplesOfLaserDynamics,LaserDynamics}. This semi-classical approximation is appropriate when spatial and polarization effects can be neglected~\cite{PrinciplesOfLaserDynamics,SemiconductorLasers,carr2006delayed}. Semiconductor lasers fall into this class, as well as CO$_{2}$, Nd:YAG, and ruby lasers~\cite{siegman86,LaserDynamics}. In their simplest form, such lasers produce stable output, with only transient relaxation oscillations above the lasing threshold\cite{siegman86,Kozyreff2023,SemiconductorLasers}. More complex dynamics can be generated by additional coupling, injection, and feedback~\cite{billings2003stochastic,kim2005scaling,schwartz1994subharmonic,SemiconductorLasers}. 

%An interesting correspondence exists between class B lasers and the canonical susceptible-infected-recovered (SIR) model for epidemic dynamics, which has been leveraged by researchers over the years to understand both\textcolor{red}{[]}. \textcolor{red}{For instance...} 

One interesting aspect of class B lasers is that, since they were derived using a population framework in their original form~\cite{Towens1999}, their dynamics is analogous to other models based on large populations. In particular, it is known that class B lasers have a direct analogy with epidemic models, such as the famous susceptible-infected-recovered (SIR) model~\cite{schwartz2004stochastic,Kozyreff2023}. In fact, one can relate the state variables of the SIR model to a class B laser, as well as physical parameters\cite{kim2005scaling}. 
%Other models that are population-based and possess mass action nonlinearities, such as molecular and biological systems, can also be found to exhibit similar complex stochastic behavior as well\cite{BLOMBERG2006133,e2011principles}. 
The result is that by studying class B lasers in detail, one can reveal a great deal about the dynamics, global manifold structure, bifurcations, and fluctuations of other large population systems~\cite{billings2004stochastic,billings2003stochastic}.

An important aspect for lasers, epidemics and large populations in general, is the role of perturbations and noise, especially when these effects excite nonlinearites and instabilities, which produce large amplitude fluctuations\cite{Agrawal1991,billings2003stochastic,PhysRevLett.107.053901,PhysRevLett.106.248102,SwitchingHindes}. In the epidemic context, recent work has shown how both demographic and parameter noise can produce unusually large and small outbreaks for systems governed according to SIR-like models\cite{PhysRevLett.128.078301,hindes2023outbreak}. For such epidemic outbreaks, infection grows stochastically in an initially susceptible population, and away from an unstable disease-free state, resulting in a probability distribution for the total fraction of a population infected, which depends on infection and recovery rates and noise characteristics. By making the analogy between infected and susceptible populations with photons and atomic carriers, we can perform a related analysis of the transient dynamics for class B lasers.    

In this work we construct an analytical theory describing how laser intensity grows away from an unstable, zero intensity equilibrium in class B lasers which are brought above their lasing threshold in the presence of random fluctuations in their electric-field dynamics. In particular, we calculate the probability (on log scale) and the most-likely dynamics that produces different observed values of the first peak in the intensity due to noise during the initial laser transient. We quantify a continuum of boundary conditions that generate different peaks and find analytical expressions for the statistics of the first intensity peak in physically relevant regimes. More broadly, our work demonstrates how to analyze the distribution of stochastic, large amplitude pulses in nonequilibrium population systems with relaxation dynamics.

Our paper is organized as follows. In Sec.\ref{sec:Setup} we introduce the model and transient dynamics for the first peak in intensity for class B lasers. In Sec.\ref{sec:Analysis} we construct a Wentzel–Kramers–Brillouin (WKB) approximation for analyzing the most-likely dynamics producing different peaks. In Sec.\ref{sec:Statistics} we study and compare theory and simulations for the statistics of the first peak. Finally, in Sec.\ref{sec:Conclusion} we offer a discussion of our results and thoughts on additional applications and generalizations.

\section{\label{sec:Setup}NOISY CLASS B LASER SYSTEMS}
To begin, consider the dynamics of class B laser systems.
In general, such systems are described by equations that govern the 
time evolution of an optical field and the number of carriers
in a laser cavity. For example, for single-mode lasers the following complex rate equations are applicable, which determine the dynamics of an electric field, $E(t)\!=\!E_{x}(t)+iE_{y}(t)$, and (normalized) carrier number, $N$ \cite{1071522,PhysRevLett.107.053901,Jin:17,photonics9020103,Paoli1988,SemiconductorLasers,LangKobayashi,PhysRevE.87.062913}:  
\begin{subequations}
\label{SemicondictorLaserSystem}
\begin{align}
&\dot{E_{x}}=k(N-1)(E_{x}-\alpha E_{y})+\sqrt{D}\xi_{x}(t),\\
&\dot{E_{y}}=k(N-1)(\alpha E_{x}+E_{y})+\sqrt{D}\xi_{y}(t),\\
 &\dot{N}=\frac{1}{\tau_{N}}(\mu-N-N|E|^{2}). 
\end{align}
\end{subequations}
Here, $k\!=\!1\big/2\tau_{p}$ where $\tau_{p}$ is the photon lifetime, $\tau_{N}$ is the carrier lifetime, $\mu$ is a pump parameter, $\alpha$ is a linewidth enhancement factor (e.g., for semiconductor lasers), and $\xi_{x}$ and $\xi_{y}$ are additive Gaussian noises with amplitudes $\sqrt{D}$ . The noise terms represent dynamical perturbations to the electric field, e.g., from spontaneous emission\cite{SemiconductorLasers,photonics9020103}. For reference, common values for parameters in typical semiconductor lasers are: $\tau_{p}\!=\!1$ ps, $\tau_{N}\!=\!1$ ns, $\alpha\!=\!3$, $D\!=\!10^{-4}$ ns$^{-2}$. When noise is neglected, Eqs.(\ref{SemicondictorLaserSystem}) are effectively two-dimensional, and depend on the laser intensity, $I\!\equiv\!E_{x}^{2}+E_{y}^{2}$, which is proportional to the photon number. Without noise, the dynamics for the intensity follows
\begin{align}
\label{NoiseFreeIntensity}
&\dot{I}=2k(N-1)I,  
\end{align}
which we refer to as the {\it deterministic} theory. %{\it mean-field}. 

If we introduce the dimensionless time $\tau\!=\!t/\tau_{p}$ and the ratio $\gamma\!=\!\tau_{p}/\tau_{N}$, we can derive a commonly analyzed form of Eqs.(\ref{SemicondictorLaserSystem}):
\begin{subequations}
\label{NormalizedLaserSystem}
\begin{align}
&\frac{dE_{x}}{d\tau}=\frac{1}{2}(N-1)(E_{x}-\alpha E_{y})+\sigma\xi_{x}(t),\\
&\frac{dE_{y}}{d\tau}=\frac{1}{2}(N-1)(\alpha E_{x}+E_{y})+\sigma\xi_{y}(t),\\
&\frac{dN}{d\tau}=\gamma(\mu-N-N|E|^{2}), 
\end{align}
\end{subequations}
or more generally a physically-inspired system describing relaxation oscillations subjected to additive Gaussian noise on a complex field with amplitude $\sigma$. Note that the noise amplitude also transforms from Eqs.(\ref{SemicondictorLaserSystem}) to Eqs.(\ref{NormalizedLaserSystem}), as $\sigma^{2}=D\gamma^{2}$. For typical semiconductor lasers with small $\gamma$ and $D$ (on $\tau_{N}$-timescales) the result is weak noise. {\it However, in this work we assume that $\sigma$ is small but non-negligible,} which will be relevant for small lasing systems and for other coupled-population systems with relaxation dynamics and relatively small system sizes. 

In terms of parameter dependence, there are two qualitatively distinct regimes. The first is below the lasing threshold $\mu\!<\!1$. In this case the laser intensity decays to zero as $\tau\!\rightarrow\!\infty$, with $I\!\rightarrow\!0$ and $N\!\rightarrow\!\mu$. On the other hand above the lasing threshold $\mu\!>\!1$: $N\!\rightarrow\!1$ and $I\!\rightarrow\!\mu\!-\!1$ as $\tau\!\rightarrow\!\infty$. In the latter case, intensity builds up in a series of giant pulses, each of which decays back down near the unstable $I\!\approx\!0$ before repeating in cyclical fashion as the stable equilibrium is eventually reached. As described, these dynamics are a canonical example of {\it relaxation oscillations}, which are sometimes called active Q-switching\cite{GiantPulse,FundamentalsOfPhotonics,Kozyreff2023}.

Most works on laser noise have considered fluctuations in special parameter regimes\cite{FokkerPlanckClassB,Paoli1988} and small fluctuations around a steady state in the form of linear-response relations\cite{SemiconductorLasers,1073018}. For example, a commonly used measure of laser intensity fluctuations is the relative intensity noise (RIN) which is the Fourier transform of $I(t)$ \cite{NAGARAJAN1999177}. 
%For example, in semiconductor lasers, the output optical power fluctuates due to electric-field noise from spontaneous emission.  
%A commonly used measure of the resulting laser intensity fluctuations is the relative intensity noise (RIN) which is the Fourier transform of $I(t)$ \cite{NAGARAJAN1999177}.
By analyzing small fluctuations of Eqs.(\ref{NormalizedLaserSystem}) near the laser steady-state, one can show that the RIN has a maximum at a predictable frequency\cite{SemiconductorLasers,1073018}. In contrast, in this work we focus on noise-induced large fluctuations in the peak intensity during the transient of Eqs.(\ref{NormalizedLaserSystem}). Such fluctuations are describable in terms of a WKB approximation in analogy with stochastic outbreaks in the SIR model\cite{hindes2023outbreak}.

\begin{figure}[h]
\center{\includegraphics[scale=0.775]{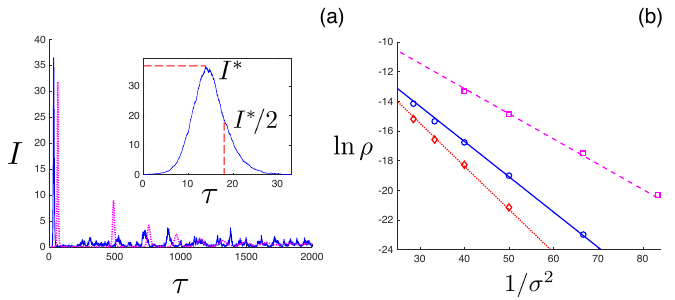}}
\caption{Transient fluctuations in Eqs.(\ref{NormalizedLaserSystem}). (a) Dynamics of the deterministic theory (dotted magenta curve) compared to a stochastic trajectory (solid blue curve). The inlet panel shows the first peak of the intensity reached by the latter. Model parameters are $\mu\!=\!1.5$, $\gamma\!=\!0.003$, $\alpha\!=\!3$, and $\sigma^{2}\!=\!0.01$. (b) The log-probability for three values of the peak intensity versus the inverse noise amplitude: $64.2\!\leq\!I^{*}\!\leq\!64.6$ (red diamonds), $61.6\!\leq\!I^{*}\!\leq\!62.0$ (blue circles), and $58.1\!\leq\!I^{*}\!\leq\!58.5$ (magenta squares). Slopes of the lines are theoretical predictions from Eqs.(\ref{HamiltonsEquations}-\ref{MomentumInitialCondition}) discussed in Sec.{\ref{sec:WKB}}. Other model parameters are $\mu\!=\!1.3$, $\gamma\!=\!0.001$, and $\alpha\!=\!3$.}
\label{fig1}
\end{figure}

The basic phenomenon that we address in this work is depicted in Fig.\ref{fig1}. In panel (a) we plot a typical time series for $I(\tau)$ in blue from Eqs.(\ref{NormalizedLaserSystem}) starting from the state $E_{x}\!=\!\eta\cos(\theta)$, $E_{y}\!=\!\eta\sin(\theta)$, $N\!=\!\mu\!>\!1$, where $\theta$ is a random phase and $\eta\!=\!10^{-6}$. From the unstable initial conditions, the laser intensity builds up to a local peak before decaying back down to near zero, and repeating. For the deterministic system with $\sigma^{2}\!\rightarrow\!0$ (plotted in magenta), subsequent local peaks decay in size as the steady state is approached on time scales $\mathcal{O}(1/\gamma)$\cite{Kozyreff2023}. 
Namely, the first peak is the largest and represents the initial burst away from the unstable equilibrium $(I\!=\!0,N\!=\!\mu)$. In the limit that the initial intensity goes to zero, the value of the peak intensity is set by $\gamma$ and $\mu$. For the stochastic system, however, a range of peak values is possible depending on how the noise drives the nonlinear dynamics. Our goal is to understand the probability distribution for the first peak in the intensity resulting from the unstable dynamics and understand the mechanics of how different peaks arise. 

To make progress, we first perform stochastic simulations, and extract the first peak, $I^{*}$. In this work, we follow the definition that the first stochastic peak corresponds to the maximum value of the intensity observed 
in a trajectory of Eqs.(\ref{NormalizedLaserSystem}) in a time window $[0,\tau_{1\!/2}]$ where $\tau_{1\!/2}$ is the time at which the intensity falls to half the maximum value observed so far. The inset panel to Fig.\ref{fig1} (a) illustrates how the first peak is extracted.

In order to get our first quantitative hold on the behavior of the first peak, we observe how the probability density for different values of $I^{*}$ depend on the noise amplitude $\sigma^{2}$. Three example series are plotted in Fig.\ref{fig1} (b). If $\sigma^{2}$ is sufficiently small, we find that the log-probabilities for each $I^{*}$ scale linearly with the inverse of the noise amplitude $1/\sigma^{2}$, with slopes that are functions of $I^{*}$. Note that each series in Fig.\ref{fig1} (b) tends to a line with a particular slope. The implied WKB scaling for the probability, $\rho(I^{*})\sim\exp(-S(I^{*})/\sigma^{2})$, is reminiscent of large deviations theory\cite{freidlin2012random,TOUCHETTE20091}. Our main goal in this work is to find the unknown function $S(I^{*})$.
% though without the typical assumption of statistical equilibrium

\section{\label{sec:WKB} WKB ANALYSIS}
Now that we have demonstrated the familiar WKB scaling form for the transient peak intensity, the next step is to analyze the dynamics of Eqs.(\ref{NormalizedLaserSystem}) in the small-noise limit, $\sigma^{2}\!\rightarrow\!0$. To do so, we make use of the following stochastic system for the intensity,
\begin{align}
\label{IntensityNoiseSystem}
&\frac{d I}{d\tau}=(N-1)I + 2\sigma\sqrt{I}\eta(\tau), %+2\sigma^{2}, 
\end{align}
where $\eta(\tau)$ is a Gaussian white-noise source. Equation (\ref{IntensityNoiseSystem}) is derived in App.\ref{sec:Appendix1}, and represents an approximation based on Eqs.(\ref{NormalizedLaserSystem}) for small $\sigma$.

Combining Eqs.(\ref{NormalizedLaserSystem}c) and Eq.(\ref{IntensityNoiseSystem}) we have the following Fokker–Planck equation\cite{StochasticProcesses,FokkerPlanck} governing the probability density for observing $I$ and $N$ at time $\tau$ (assuming It\^{o} calculus), 
\begin{align}
\label{FokkerPlanck}
\frac{\partial \rho}{\partial \tau} = &-\frac{\partial}{\partial N}\Big[\gamma(\mu-N-NI)\rho\Big] \nonumber \\
%&-\frac{\partial}{\partial I}\Big[((N-1)I+2\sigma^{2})\rho\Big] +\frac{1}{2}\frac{\partial^{2}}{\partial I^{2}}[4\sigma^{2}I].
&-\frac{\partial}{\partial I}\Big[(N-1)I\rho\Big] +\frac{1}{2}\frac{\partial^{2}}{\partial I^{2}}[4\sigma^{2}I]. 
\end{align}
Based on our observations from Fig.\ref{fig1}(b), we substitute the ansatz, 
$\rho\sim\exp[-S(I,N,\tau)/\sigma^{2}]$, into Eq.(\ref{FokkerPlanck}) and collect terms at $\mathcal{O}(1/\sigma^{2})$, which dominate as $\sigma^{2}\!\rightarrow\!0$. The result is a Hamilton-Jacobi equation, $\partial S/\partial \tau +H(I,N,\partial S/\partial I,\partial S/\partial N)\!=\!0$\cite{freidlin2012random,doi:10.1137/17M1142028,Assaf_2017,Dykman1994}, where   
\begin{align}
\label{Hamiltonian}
H=\gamma(\mu-N-NI)\lambda_{N}+(N-1)I\lambda_{I}+2I\lambda_{I}^{2}.  
\end{align}
In Equation (\ref{Hamiltonian}), $\lambda_{I}\!\equiv\!\partial S/\partial I$ and $\lambda_{N}\!\equiv\!\partial S/\partial N$ are the conjugate momenta for intensity and carrier number, respectively. We point out that even with the known transformation between class B lasers and the SIR model\cite{kim2005scaling,Kozyreff2023}, the Hamiltonian Eq.(\ref{Hamiltonian}) is different from other works, because of the distinct noise contribution from the electric-field dynamics and the singular parameter $\gamma$.

Of interest to us are solutions where initially $N\!\rightarrow\!\mu$ and $I\!\rightarrow\!0$, and hence $H\!\rightarrow\!0$ in Eq.(\ref{Hamiltonian}). Because, the Hamiltonian $(H)$ is not an explicit function of time, $H\!=\!0$ is {\it conserved} in time. This means that we need to solve Hamilton's equations\cite{freidlin2012random,doi:10.1137/17M1142028,Assaf_2017,Dykman1994}:  
\begin{subequations}
\label{HamiltonsEquations}
\begin{align}
\frac{dI}{d\tau}&=(N-1)I +4I\lambda_{I},\\
\frac{dN}{d\tau}&=\gamma(\mu-N-NI),\\
-\frac{d\lambda_{I}}{d\tau}&=-\gamma N \lambda_{N} +(N-1)\lambda_{I} +2\lambda_{I}^2\\ 
-\frac{d\lambda_{N}}{d\tau}&=-\gamma\lambda_{N}(1+I)+I\lambda_{I}, 
\end{align}
\end{subequations}
with zero energy $H\!=\!0$. Once we have such solutions, the probability exponent (called the {\it action}) for large fluctuations in laser intensity $I$ and carrier number $N$ can be computed according to the integral\cite{freidlin2012random,doi:10.1137/17M1142028,Assaf_2017,Dykman1994}, just as in analytical mechanics: 
\begin{align}
\label{ActionIntegral}
S=\int\!\lambda_{I}dI + \int\!\lambda_{N}dN.   
\end{align}

\subsection{\label{sec:NumericalSolutions} Solutions to Hamilton's equations}
Next, we turn to constructing the relevant solutions of Hamilton's Eqs.(\ref{HamiltonsEquations}) for predicting the peak-intensity distribution. For stochastic outbreaks in SIR models without births, deaths, and reinfection, one enforces fixed-point boundary conditions on the momenta, leading to a continuum of solutions with $H\!=\!0$, parameterized by a free momentum initial condition\cite{PhysRevLett.128.078301,hindes2023outbreak}. Each initial condition produces a different value of the epidemic size with a corresponding action. We follow a similar approach for class B lasers by analogy.

To find the initial conditions, we set $H\!=\!0$ at $\tau\!=\!0$ in Eq.(\ref{Hamiltonian}), and solve for the momenta.  Let us adopt the notation $\lambda_{I}(\tau\!=\!0)\!\equiv\!\lambda_{I}^{0}$, and similarly for all phase-space variables. Recall that $N^{0}\!=\!\mu$ and $I^{0}\!=\!0$. As already pointed out, if we plug $N^{0}$ and $I^{0}$ into Eq.(\ref{Hamiltonian}), we get $H^{0}\!=\!0$ independent of the momenta. However, if we enforce $H^{0}\!=\!0$ independent of $I^{0}$, we get a one-parameter family of zero-energy initial conditions:
\begin{align}
\label{MomentumInitialCondition}
&\lambda_{N}^{0}=\frac{(\mu-1)\lambda_{I}^{0}+2{\lambda_{I}^{0}}^{2}}{\mu\gamma},\nonumber \\
&I^{0}=0, \;\;\text{and}\;\; N^{0}=\mu,  
\end{align}
parameterized by the laser intensity momentum $\lambda_{I}^{0}$. %For reference, in the epidemic systems\textcolor{red}{[]}, the momenta initial conditions produce $H^{0}\!=\!0$ independent of the fraction of the population infected. 

For the class B laser model studied, the initial conditions Eq.(\ref{MomentumInitialCondition}) imply that  $d\lambda_{I}/d\tau\!=\!0$ Eq.(\ref{HamiltonsEquations}c) at $\tau\!=\!0$, while $d\lambda_{N}/d\tau\!\neq\!0$ in Eq.(\ref{HamiltonsEquations}d) for $\gamma\!\neq\!0$. In contrast, in the SIR model without births, deaths, and reinfection the conjugate momenta all have fixed-point boundary conditions\cite{PhysRevLett.128.078301,hindes2023outbreak}. In this way the laser system has more general outbreak solutions because $\gamma\!\neq
\!0$. Recall that the finite $\gamma$ is responsible for the relaxation oscillations, which are not present in epidemic models without long timescale demographic dynamics.

Now that we know the initial conditions, we need to construct solution paths forward in time from Eq.(\ref{MomentumInitialCondition}) and Eqs.(\ref{HamiltonsEquations}), find the first peak in the intensity for every $\lambda_{I}^{0}$ where $I(\tau)$ has its first local maximum ($I^{*}$), and substitute trajectories to the peak intensity $I^{*}$ into Eq.(\ref{ActionIntegral}) to find the action. In general, we must do this numerically for arbitrary $\mu$ and $\gamma$ by sweeping over values of $\lambda_{I}^{0}$. 
%\IBS{I do not think explaining details of ODE solutions are really needed. Maybe just state the ICs. Given that, anyone should be able to simulate what you did.}
%A straightforward algorithm is to: (1) Pick a $\lambda_{I}^{0}$. (2) Integrate Eqs.(\ref{HamiltonsEquations}) with a small random perturbation (e.g., $10^{-10}$) on the initial condition Eq.(\ref{MomentumInitialCondition}). Equations (\ref{HamiltonsEquations}) are integrated until $I(\tau)$ reaches a {\it local} maximum, $I^{*}$, i.e., when 
%$dI/d\tau\!=\!-10^{-4}$. 
%$dI/d\tau$ is less than some infinitesimal negative threshold. (3) Compute the integral Eq.(\ref{ActionIntegral}) to $I^{*}$. (4) Repeat for a different $\lambda_{I}^{0}$, generating a different $I^{*}$ with a different action\footnote{Using small time steps, $d\tau\!=\!10^{-4}$, we can achieve $H\!=\!0$ to double precision with, e.g., Runga-Kutta integration.}. 
The slopes of the lines in Fig.\ref{fig1} (b) were computed using numerical ODE solvers with small random perturbations to the initial conditions Eq.(\ref{MomentumInitialCondition}). For each of the three example values of $I^{*}$ plotted, we see good agreement between stochastic simulations and numerical solutions for the action. We return to probability density comparisons between stochastic simulations and the WKB solution in more detail in Sec.\ref{sec:Statistics}.

\subsubsection{Example solutions}
\begin{figure}[t]
\center{\includegraphics[scale=0.77]{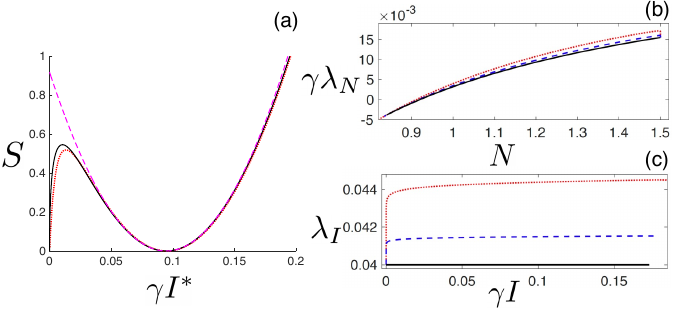}}
\caption{WKB theory illustration for the first peak in the intensity. (a) Action versus the peak intensity for $\gamma\!=\!0.003$ (dotted red), the $\gamma\!\rightarrow\!0$ approximation Eqs.(\ref{Istar}-\ref{Sstar}) (solid black), and the quadratic approximation for $\gamma\!\rightarrow\!0$  Eqs.(\ref{Squad}-\ref{Ilin}) (dotted magenta). (b) Example conjugate momentum trajectories for the carrier number versus the carrier number for $\gamma\!=\!0.003$ (dotted red), $\gamma\!=\!0.001$ (dashed blue), and the $\gamma\!\rightarrow\!0$ approximation Eq.(\ref{MomentumSmallGamma}) (solid black). (c) Example conjugate momentum trajectories for the laser intensity versus the laser intensity for $\gamma\!=\!0.003$ (dotted red), $\gamma\!=\!0.001$ (dashed blue), and the $\gamma\!\rightarrow\!0$ approximation (solid black). For all panels $\mu\!=\!1.5$ and $\alpha\!=\!3$.}
\label{fig2}
\end{figure}

Before continuing our analysis let us examine the basic structure of WKB solutions. First, by sweeping $\lambda_{I}^{0}$ in Eq.(\ref{MomentumInitialCondition}), we can generate the full action for a given $\mu$ and $\gamma$. An example is plotted in Fig.\ref{fig2} (a) with a dotted red curve. The two additional curves plotted in Fig.\ref{fig2} (a) are analytical approximations to the red curve and are discussed in detail in Secs.\ref{sec:Analysis}-\ref{sec:Statistics}. Similar to the case of stochastic outbreaks\cite{hindes2023outbreak}, the action has a cubic structure, where it takes on a local minimum $S\!=\!0$ at the deterministic value ($\gamma I^{*}\!\approx\!0.1$ in the example shown) and for infinitesimally small peak intensities ($\gamma I^{*}\!\rightarrow\!0$). The fact that both the deterministic and zero-intensity solutions have vanishing actions for the first peak is analogous to the ``boom or bust" dynamics common in stochastic population systems\cite{Strayer2017}. In between the two $S\!=\!0$ cases, the action has a single maximum, and then monotonically increases above the deterministic value. Given the non-monotonic action with multiple regimes, we can see that the distribution of peak intensities is predicted to be non-Gaussian in general.

Second, in addition to the action, we can look at the phase-space trajectories. Several examples are shown in Fig.\ref{fig2}, where we plot carrier number momenta and intensity momenta versus their respective conjugates in panels (b) and (c). Within each panel, solutions for three different values of $\gamma$ are shown. Most importantly, as $\gamma\!\rightarrow\!0$ (the black curves), the momenta converge to a limiting form, where $\lambda_{I}$ tends to a {\it constant} in time and $\gamma \lambda_{N}$ tends to some function of $N$. The small-$\gamma$ limit is relevant for typical class B laser systems, and is analyzed further in Sec.\ref{sec:Analysis}.

\begin{figure}[t]
\center{\includegraphics[scale=0.748]{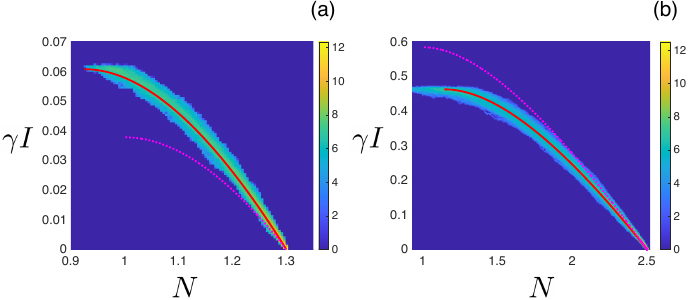}}
\caption{Probability heat maps of stochastic trajectories from Eqs.(\ref{NormalizedLaserSystem}) that result in peak intensity above (a) and below (b) the deterministic theory. The color maps for the probability are given on log scale from $500$ simulations. Curves represent: theoretical predictions (solid red lines) and deterministic theory (dotted magenta lines). (a) Trajectories resulting in $61.6\!\leq\!I^{*}\!\leq\!62.0$ with $\mu\!=\!1.3$ and $\sigma^{2}\!=\!0.015$. (b) Trajectories resulting in $466\!\leq\!I^{*}\!\leq\!470$  with $\mu\!=\!2.5$ and $\sigma^{2}\!=\!0.2$. For both panels $\gamma\!=\!0.001$ and $\alpha\!=\!3$. Theoretical predictions come from solving Eqs.(\ref{HamiltonsEquations}) given the initial conditions Eq.(\ref{MomentumInitialCondition}).}
\label{fig3}
\end{figure}

Finally, in general, solutions to Eqs.(\ref{HamiltonsEquations}) result in minimized actions, given the boundary conditions, and therefore represent locally maximal-probability solutions, or most-likely paths for Eqs.(\ref{NormalizedLaserSystem}) in the small-noise limit\cite{freidlin2012random,doi:10.1137/17M1142028,Assaf_2017,Dykman1994}. Therefore, we expect the stochastic laser system to produce intensity and carrier number trajectories that exist within a narrow tube of probability, centered around solutions to Eqs.(\ref{HamiltonsEquations}). To illustrate, we show two examples in Fig.\ref{fig3}. Probability heat maps of the intensity versus carrier number from $500$ stochastic simulations of Eqs.(\ref{NormalizedLaserSystem}) for $I^{*}$ above (a) and below (b) the deterministic value are plotted. The solution of Eqs.(\ref{HamiltonsEquations}) is plotted in red along side the deterministic solution in magenta. As we expect, the WKB solution passes through the relatively small cloud of stochastic trajectories, and is well separated from the deterministic theory. We can also see that, in general, the structure of the dynamics to the first intensity peak seems relatively simple and monotonic in, e.g., the carrier number. Combining this pattern with the apparent limiting form as $\gamma\!\rightarrow\!0$, we can gain further analytical understanding of Eqs.(\ref{HamiltonsEquations}).
%\IBS{I know we chatted about it, but a heat map showing the analytical approximation goes through high probability (density) regions of fig. 3 would be nice. But it might be too much computation for the result.}
\subsection{\label{sec:Analysis} Analytical solution for $\gamma\!\rightarrow\!0$}
Now, we return to analyzing the WKB theory. In particular, for typical class B laser systems $\gamma$ is usually on the order of $10^{-3}$\cite{SemiconductorLasers,LaserDynamics,Kozyreff2023}. Motivated by this small magnitude and the comparison plots of Fig.\ref{fig2} (b-c), we construct analytical solutions
for Eqs.(\ref{HamiltonsEquations}-\ref{ActionIntegral}) in the limit $\gamma\!\rightarrow\!0$, which serve as finite-$\gamma$ approximations and useful conceptual guides. First, we start with the scaling-form suggested in Fig.\ref{fig2} (b-c). Namely, as $\gamma\!\rightarrow\!0$: $I\!\rightarrow\!a/\gamma$ and $\lambda_{N}\!\rightarrow\!p_{N}/\gamma$, where $a(\tau)$ and $p_{N}(\tau)$ are functions independent of $\gamma$. We can substitute this ansatz into Eq.(\ref{Hamiltonian}), set $H\!=\!0$, and take $\gamma\!\rightarrow\!0$. The result is
\begin{align}
\label{MomentumSmallGamma}
p_{N}= \frac{(N-1)\lambda_{I}+2\lambda_{I}^{2}}{N}. 
\end{align}
Importantly, Eq.(\ref{MomentumInitialCondition}) implies that $d\lambda_{I}/d\tau\!=\!0$ in general from Eqs.(\ref{HamiltonsEquations} c). Therefore, as $\gamma\!\rightarrow\!0$ a conserved zero-energy implies a {\it conserved momentum}. The same conservation of a continuous momentum appears in epidemic outbreaks in the SIR model with demographic noise\cite{PhysRevLett.128.078301}.

Because $\lambda_{I}$ is conserved as $\gamma\!\rightarrow\!0$, $\lambda_{N}$ only depends on $N$ according to Eq.(\ref{MomentumSmallGamma}) along a trajectory. Our next step is to find a similar function, $a(N)$. By substituting $I\!=\!a/\gamma$ into Eqs.(\ref{HamiltonsEquations} a-b), and taking their ratio in the limit $\gamma\!\rightarrow\!0$, we can write: 
\begin{align}
\label{dadN}
\frac{da}{dN}=-\frac{(N-1)+4\lambda_{I}^{0}}{N}.   
\end{align}
Solving Eq.(\ref{dadN}) by separating variables and integrating both sides from the equilibrium $(a\!=\!0,N\!=\!\mu)$, we find
\begin{align}
\label{a}
a(N)=\mu-N+(1-4\lambda_{I}^{0})\ln(N/\mu).  
\end{align}

At this point, since all of the phase-space variables are explicit functions of $N$ and the conserved momentum $\lambda_{I}^{0}$, we are in a position to calculate the peak laser intensity and its associated action. To find the peak intensity $I^{*}\!=\!a^{*}/\gamma$, we set $da/dN\!=\!0$ and solve for the corresponding carrier number, $N^{*}$, or
\begin{align}
\label{Nstar}
N^{*}=1-4\lambda_{I}^{0}. 
\end{align}
Substituting Eq.(\ref{Nstar}) into Eq.(\ref{a}) results in 
\begin{align}
\label{Istar}
I^{*}(N^{*})= \frac{\mu-N^{*}+N^{*}\ln(N^{*}/\mu)}{\gamma}.  
\end{align}

%\IBS{Can I see your computation for the action? }
Finally, we integrate Eq.(\ref{ActionIntegral}) to the point of maximum intensity $(I^{*},N^{*})$, which is expressed explicitly in terms of $N^{*}$: 
\begin{align}
\label{Sstar}
S^{*}(N^{*})= \frac{(1-N^{*})(N^{*}-1)\ln(N^{*}/\mu)}{8\gamma}. 
\end{align}
Equations (\ref{Istar}-\ref{Sstar}) give the complete solution for the log-probability of $I^{*}$ as $\gamma\!\rightarrow\!0$ and $\sigma^{2}\rightarrow\!0$. To generate $S^{*}(I^{*})$ one simply varies $N^{*}$ over $[1,\infty)$ in Eqs.(\ref{Istar}-\ref{Sstar}). 

We now return to Fig.\ref{fig2}, where in each of the panels the analytical approximation for small $\gamma$ is compared to numerical solutions of Hamilton's equations described in Sec.\ref{sec:NumericalSolutions}. For panels (b) and (c), we pick a particular initial condition, $\lambda_{I}^{0}\!=\!0.04$, and compare Eq.(\ref{MomentumSmallGamma}) and the conserved $\lambda_{I}$ assumption, respectively, to two trajectories with representative values of $\gamma$ for a class B laser. Each of the plots demonstrates a convergence to the analytical theory as $\gamma\!\rightarrow\!0$. In panel (a), the numerical solution for the action Eq.(\ref{ActionIntegral}) when $\gamma\!=\!0.003$ is compared to Eq.(\ref{Istar}-\ref{Sstar}). Here too, we find good agreement, where the analytical theory captures the cubic structure of the action, with fractional error that is $\mathcal{O}(\gamma)$.

In addition to the action, another useful aspect of the small-$\gamma$ approximation is the effective update to the deterministic Eq.(\ref{NoiseFreeIntensity}) as a result of the noise. Namely, since $\lambda_{I}^{0}$ is approximately conserved, the intensity dynamics becomes 
\begin{align}
\label{MFupdate}
\frac{dI}{d\tau}&\approx(N-1)I +4\lambda_{I}^{0}I,
\end{align}
or the laser deterministic theory plus an additional linear growth/damping term depending on the value of the conserved fluctuational momentum.

\section{\label{sec:Statistics} STATISTICS OF THE PEAK INTENSITY}
In this section we turn to the statistics of the transient fluctuations of Eqs.(\ref{NormalizedLaserSystem}). First, we consider typical fluctuations around the deterministic solution, recalling that the deterministic solution corresponds to the noiseless limit, $\sigma^{2}\!\rightarrow\!0$. In particular, we are interested in how the mean and variance of small fluctuations away from the deterministic 
peak ($I_{\text{D}}^{*}$) behave for finite $\sigma^{2}$ and $\gamma$. Unfortunately, shifts in the mean laser intensity are not accounted for by the WKB solution for finite $\sigma^{2}$, since they reside in sub-exponential corrections to the probability distribution. In this work, we extract the approximate scaling with stochastic simulations. 

\begin{figure}[h]
\center{\includegraphics[scale=0.702]{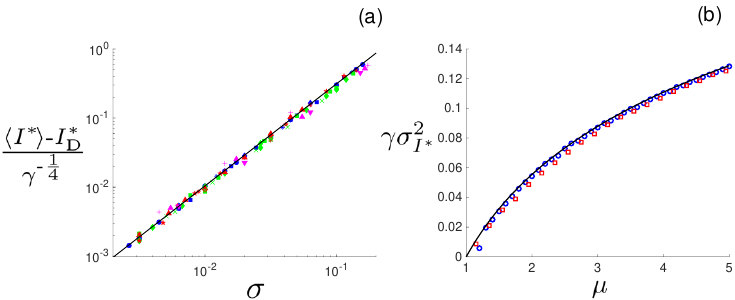}}
\caption{Statistical moments of the first peak of the intensity that result in large peaks. (a) Shift in the mean as a function of noise intensity. Different colors represent: $\mu\!=\!1.2$ (red), $\mu\!=\!1.3$ (blue), $\mu\!=\!1.5$ (green), and $\mu\!=\!1.7$ (magenta), while different symbols represent values of $\gamma$ ranging uniformly from $5*10^{-5}$ to $5*10^{-2}$. The line of best-fit for the scaling is plotted in black. (b) Variance as a function of pump parameter. Stochastic simulation results are shown for $\gamma\!=\!0.001$ (blue circles) and $\gamma\!=\!0.05$ (red squares). Theoretical predictions from Eq.(\ref{SmallGammaLargePeakVariance}) are shown with a black curve. Other model parameters are $\alpha\!=\!3$, and $\sigma^{2}=0.02$.}
\label{fig4}
\end{figure}

Since there is a significant probability for the first transient peak to result in an infinitesimal intensity, as pointed out in Sec.\ref{sec:NumericalSolutions}, and we are interested in small fluctuations relative to the deterministic peak ($I_{\text{D}}^{*}$), we perform stochastic simulations and average the results that produce peak intensities which are above a small threshold, e.g, $I^{*}\!>\!0.05/\gamma$\footnote{For computing the mean and variance of the large peaks in Fig.\ref{fig4} we use $1000$ stochastic simulations that result in peak intensities above the value $0.05/\gamma$.}. 
In the epidemic context, such extensive outbreaks are called ``large outbreaks" \cite{PhysRevLett.128.078301,hindes2023outbreak}. In analogy, we call them ``large” intensity peaks, here. %\IBS{I would be careful with carrying the analogy too far, since a referee might say if you just show the general equations of motion for both, the problem is done.}

In Fig.\ref{fig4} (a) we plot the shift in the mean-intensity for large peaks relative to the deterministic theory as a function of noise amplitude. Several orders of magnitude are sampled in $\gamma$ and $\sigma^{2}$, and the results are shown for several values of the pump parameter $(\mu)$. The main trend is clear that we observe a power-law scaling of the form,
\begin{align}
\label{scaling}
\gamma\left<I^{*}\right>-\gamma I_{\text{D}}^{*} \sim C(\mu)(\sigma^{2})^{\alpha}(\gamma)^{\beta}.
\end{align}
By taking the natural log of both sides of Eq.(\ref{scaling}), we can find the line of best-fit for the scaling parameters using linear least squares. The values from the plotted data in Fig.\ref{fig4} (a) are: $C(\mu)\approx 9$, $\alpha\approx0.74$, and $\alpha\approx0.73$. We note that the latter two exponents are quite close to the rational $3/4$, indicative of sublinear scaling for the average shift.  

Next, we consider the variance of the large peaks. Unlike the mean, the variance of the large peaks can be extracted within the WKB approximation. In particular, we can find analytical expressions for small $\gamma$ by expanding Eqs.(\ref{Istar}-\ref{Sstar}) in small deviations around the deterministic solution. Specifically, let us assume $N^{*}
\!=\!1+\epsilon$ with $\epsilon\!\ll\!1$. Equation (\ref{Sstar}) is quadratic at lowest order in $\epsilon$, 
\begin{align}
\label{Squad}
S^{*}(\epsilon)=\frac{\ln(\mu)\epsilon^{2}}{8\gamma}+\mathcal{O}(\epsilon^{3}). \end{align}
To find the variance in $I^{*}$ we want to express $\epsilon$ in terms of $I^{*}\!-\!\bar{I}^{*}$, where $\bar{I}^{*}\!=\!(\mu-1-\ln(\mu))/\gamma$ is the deterministic solution when $\gamma\!\rightarrow\!0$. From Eq.(\ref{Istar}), it is easy to show that $I^{*}-\bar{I}^{*}$ is linear at lowest order in $\epsilon$,
\begin{align}
\label{Ilin}
I^{*}-\bar{I}^{*}=\frac{\ln{\mu}}{\gamma}\epsilon+\mathcal{O}(\epsilon^{2}).
\end{align}
Combining Eqs.(\ref{Squad}-\ref{Ilin}), and recalling that $\rho(I^{*})\sim\exp(-S(I^{*})/\sigma^{2})$, we find a Gaussian approximation for the probability distribution of $I^{*}$ around $\bar{I}^{*}$ for small $\gamma$ with variance 
\begin{align}
\label{SmallGammaLargePeakVariance}
\sigma^{2}_{I^{*}}=\frac{4\ln(\mu)\sigma^{2}}{\gamma}. 
\end{align}
We point out that the logarithmic dependence on the pump rate in Eq.(\ref{SmallGammaLargePeakVariance}) entails only a slow change in the characteristic variation in the large peaks as a laser is driven increasingly above threshold.

Comparisons between Eq.(\ref{SmallGammaLargePeakVariance}) and stochastic simulations are shown in Fig.\ref{fig4} (b), where we plot the variance in the large peak intensity versus the laser pump parameter for two values of $\gamma$. We can see that the agreement between theory and simulation is quite good -- even for the red-square series which has a relatively large ratio for the photon-to-carrier lifetimes, $\gamma\!=\!0.05$. Theses results suggest that a Gaussian approximation is accurate for capturing relatively small fluctuations around the deterministic solution when the noise amplitude is weak.

\begin{figure}[t]
\center{\includegraphics[scale=0.765]{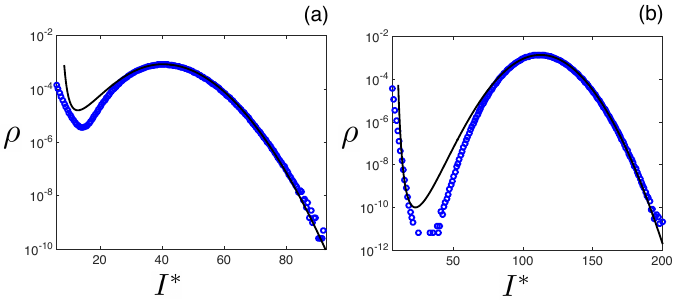}}
\caption{Probability density for the first peak in the intensity during the transient relaxation of Eqs.(\ref{NormalizedLaserSystem}). Stochastic simulation results are shown with blue circles while theoretical predictions are shown with a black curve from solving Eqs.(\ref{HamiltonsEquations}-\ref{MomentumInitialCondition}) and adding the measured shift for finite $\sigma^{2}$, Eq.(\ref{scaling}). (a) $\mu\!=\!1.5$ and $\sigma^{2}\!=\!0.15$. (b) $\mu\!=\!2.0$ and $\sigma^{2}\!=\!0.2$. For both panels $\gamma\!=\!0.003$ and $\alpha\!=\!3$.}
\label{fig5}
\end{figure}

On the other hand, large fluctuations are possible for which we must consider the full distribution, beyond the Gaussian approximation. In order to sample probability densities well into the non-Gaussian tails, we must use fairly large noise amplitudes $\sigma^{2}\!\sim\!\mathcal{O}(10^{-1})$. As a consequence, we add the measured shift in the peak Eq.(\ref{scaling}), otherwise the WKB solution and finite noise mean values for the large peaks are offset by roughly $5\%$\footnote{
Namely, for every value of $I^{*}$ computed with Eqs.(\ref{HamiltonsEquations}) and Eq.(\ref{MomentumInitialCondition}), we add $C(\mu)(\sigma^{2})^{\alpha}(\gamma)^{\beta}
$ with the least-squares values measured in Fig.\ref{fig4}}. Two example comparisons between stochastic simulations and numerical solutions of Eqs.(\ref{HamiltonsEquations}-\ref{ActionIntegral}) are shown in Fig.\ref{fig5}, for two values of the pump parameter. The simulation-based probability densities are computed from $\sim\!10^{11}$ stochastic realizations. Overall, we find good agreement for probability-density estimates above the mean, and fair agreement below the mean. The latter are generally still accurate to within an order of magnitude. Moreover, the WKB result captures the cubic qualitative structure of the distribution on log scale predicted in Sec.\ref{sec:NumericalSolutions}, and is very similarly to the case of epidemic outbreaks in the presence of stochasticity. This combination of theory and simulations demonstrates that the tails of the peak intensity distribution for class B lasers are highly non-Gaussian, yet describable by optimal noise fluctuations that drive transient growth in laser intensity away from an unstable equilibrium.   

%\section{\label{sec:Conclusion} CONCLUSIONS}
\section{\label{sec:Conclusion} SUMMARY AND
CONCLUSIONS}
Lasers exhibit complex dynamics that emerges from 
inherent nonlinearities, instabilities, and various forms of coupling. A basic problem in laser physics is to understand the response to perturbations and noise, especially when such effects result in large amplitude fluctuations. In this work, we focus on the noisy transient dynamics of single-mode class B lasers, brought suddenly above threshold, in which intensity grows away from an unstable equilibrium to a local maximum, or first peak. The first peak can take on a range of values because of noise in the electric-field dynamics, e.g., from spontaneous emission of photons. This process is closely aligned with outbreak dynamics in the Susceptible-Infected-Recovered (SIR) model of epidemics, where different noise sources can produce unusually large and small outbreaks compared to deterministic assumptions\cite{PhysRevLett.128.078301,hindes2023outbreak}. 

Recent work on the distribution of such stochastic outbreaks
starting in an unstable, fully susceptible population,
has shown that in the small-noise limit, the probability 
is determined by a one-parameter family
of Hamiltonian paths, describeable using a WKB approximation\cite{PhysRevLett.128.078301,hindes2023outbreak}. In this work, 
we constructed a similar but more general family of solutions 
for class B lasers where each first-peak in the intensity has
unique fluctuational-momentum initial conditions, 
found an analytical solution for the probability on log-scale of the first peak
in the limit where the ratio of photon-to-carrier lifetimes
is small, and calculated the small-fluctuation variance of
the first-intensity peak around the deterministic theory.
More generally, our work shows how to construct outbreak distributions in population systems with nonequilibrium and relaxation dynamics away from an unstable state.  

From a broad perspective, our results can form the basis for analyzing similar fluctuations in non-equilibrium and unstable transients in a wider array of laser problems. Examples include peak-intensity fluctuations in multi-mode lasers, class B lasers with different noise sources and characteristics, delayed feedback lasers, and coupled laser arrays. In addition, the theory can be augmented with state-dependent controls whereby unusual fluctuations are rendered exponentially more (or less) likely to occur~\cite{schwartz2004dynamical,PhysRevLett.117.028302}. This may shed light on ways to leverage noise in simple laser systems in order to produce large intensity pulses. Lastly, since the SIR model has been shown to underlie many processes in physical, biological, and social sciences, the approach presented will likely find application in many other transient fluctuation problems in coupled population systems.

\section*{ACKNOWLEDGEMENTS}
JH and IBS were supported by the U.S. Naval Research Laboratory funding
(N0001419WX00055), and the Office of Naval Research (N0001419WX01166) and
(N0001419WX01322).

\begin{appendix} 
\section{\label{sec:Appendix1} Stochastic system for intensity}
From Eqs.(\ref{NormalizedLaserSystem}) we want to find an approximate stochastic differential equation for the laser intensity that makes the WKB analysis of Sec.\ref{sec:WKB} simpler. To do so, we start with the Milstein method, which gives a discrete-time approximation for a stochastic differential equation with fixed time steps equal to $\Delta\tau$. The approximation is accurate to $\mathcal{O}(\Delta \tau)$\cite{StochasticProcesses}. In fact, since the noise terms in Eqs.(\ref{NormalizedLaserSystem}) are additive Gaussian, the Milstein method reduces to the simple Euler-Maruyama formula\cite{StochasticProcesses}, or:
\begin{subequations}
\label{dEs}
\begin{align}
&\Delta E_{x}= \frac{\Delta \tau}{2}(N(\tau)-1)(E_{x}(\tau)-\alpha E_{y}(\tau))+\sigma\sqrt{\Delta \tau}G_{x}(0,1)\\
&\Delta E_{y}= \frac{\Delta \tau}{2}(N(\tau)-1)(\alpha E_{x}(\tau)+E_{y}(\tau))+\sigma\sqrt{\Delta \tau}G_{y}(0,1), 
\end{align}
\end{subequations}
where $\Delta E_{x}\!\equiv\!E_{x}(\tau+\Delta\tau)-E_{x}(\tau)$ and $\Delta E_{y}\!\equiv\!E_{y}(\tau+\Delta\tau)-E_{y}(\tau)$, and where $G_{x}(0,1)$ and $G_{y}(0,1)$ represent stochastic contributions drawn from two independent Gaussian distributions with zero mean and unit variance. %\IBS{I think you are missing a $\Delta \tau$ in the deterministic part. Please check.}

Next, if we write the electric field components as $E_{x}=\sqrt{I}\cos(\psi)$ and $E_{y}=\sqrt{I}\sin(\psi)$, the change in the intensity over a time step $\Delta\tau$, $\Delta I\!\equiv\!I(\tau+d\tau)-I(\tau)$, is
\begin{align}
\label{DeltaI1}
\Delta I =&I(\tau)(N(\tau)-1)\Delta \tau\;\;+\nonumber \\
&2\sigma\sqrt{\Delta \tau}\sqrt{I(\tau)}\big(\cos(\psi)G_{x}(0,1)\!+\!\sin(\psi)G_{y}(0,1)\big)\;+ \nonumber \\
&\Delta E_{x}^{2}+ \Delta E_{y}^{2}. 
\end{align}
The second term in Eq.(\ref{DeltaI1}) proportional to $\sqrt{\Delta \tau}$, is the sum of independent Gaussians with zero mean and variance $\cos^{2}(\psi)\!+\!
\sin(\psi)^{2}\!=\!1$. Hence, we can write
\begin{align}
\label{DeltaI2}
\Delta I =&I(\tau)(N(\tau)-1)\Delta \tau+2\sigma\sqrt{\Delta \tau}\sqrt{I(\tau)}G_{I}(0,1)\;+\nonumber\\
&\Delta E_{x}^{2}+ \Delta E_{y}^{2}. 
\end{align}
%\IBS{Same issue for the intensity equation}

The final contribution to $\Delta I$  in Eq.(\ref{DeltaI2}) is
\begin{align}
\label{DeltaI3}
\Delta E_{x}^{2}+ \Delta E_{y}^{2}= \sigma^{2}\Delta\tau\big(G_{x}(0,1)^{2}+G_{y}(0,1)^{2}\big)+\mathcal{O}(\Delta\tau^{3/2}). 
\end{align}
Namely, at lowest order in $\Delta\tau$, Eq.(\ref{DeltaI3}) is a chi-squared-$2$ distributed random variable with mean equal to $2\sigma^{2}\Delta\tau$. Since the contribution is $\mathcal{O}(\sigma^{2})$,
we neglect it in our approximation of Eq.(\ref{DeltaI2}) for small $\sigma$. Doing so as $\Delta\tau\!\rightarrow\!0$, gives Eq.(\ref{IntensityNoiseSystem}) in the main text. We note that if one approximates  
Eq.(\ref{DeltaI3}) by the mean contribution at $\mathcal{O}(\Delta\tau)$, $2\sigma^{2}\Delta\tau$, the WKB analysis of Sec.\ref{sec:WKB} remains unchanged.  

%In order to simplify the analysis we approximate Eq.(\ref{DeltaI3}) by its mean value. Doing so, we arrive at the approximation Eq.(\ref{IntensityNoiseSystem}) in the main text, as $\Delta\tau\!\rightarrow\!0$.

%We note that for the LDT analysis, the positive term $2\sigma^{2}$ in Eq.(\ref{IntensityNoiseSystem}), 
%found by approximating Eq.(\ref{DeltaI3}) by its mean value, does not contribute when we take the singular limit ($\sigma^{2}\rightarrow\!0$) of Eq.(\ref{FokkerPlanck}). Namely, our approximation serves as a convenient place-holder, only, for higher order $\sigma$-corrections to the intensity dynamics.

\end{appendix} 

\bibliographystyle{unsrt}
\bibliography{lasers,laser_apps}

\end{document}